\documentclass[10pt,conference]{IEEEtran}
\usepackage[cmex10]{amsmath}
\usepackage[utf8]{inputenc}
\usepackage{multicol}
\usepackage{graphicx}
\usepackage{textcomp}
\usepackage{pdfpages}
\usepackage{pdflscape}
\usepackage{mdwmath}
\usepackage{mdwtab}
\usepackage{subfig}
\usepackage{float}
\usepackage{url}
\usepackage{enumerate}
\usepackage{breqn}
\usepackage{amssymb}
 
\IEEEoverridecommandlockouts

\begin{document}
\title{A Heuristic Algorithm for Network Optimization of OTN over DWDM Network}

\author{
\IEEEauthorblockN{Govardan~C.$^*$, Sri~Krishna Chaitanya K., Krishna Kumar Naik~B.,\\ Shreesha Rao~D.~S.,
Jagadeesh C.,  Gowrishankar~R. and Siva Sankara Sai~S.}
\IEEEauthorblockA{Department of Physics, Sri Sathya Sai Institute of Higher Learning,\\ Prasanthinilayam, Andhra Pradesh \hspace{1 cm}  
}
%\IEEEcompsocitemizethanks{\IEEEcompsocthanksitem$^*$Govardan~C is currently with Infinera India Pvt. Ltd.\protect\\
%E-mail: gchandrababu@infinera.com}

\IEEEcompsocitemizethanks{\IEEEcompsocthanksitem$^*$E-mail:gchandrababu@infinera.com}
\and

\IEEEauthorblockN{Prabhat Behere$^\dagger$ \\ and Bhyri Sai Kishore$^\ddag$ }

\IEEEauthorblockA{ Cisco Systems Pvt. Ltd. \\ Bengaluru, Karnataka}
\IEEEcompsocitemizethanks{ \IEEEcompsocthanksitem $^\dagger$E-mail:pbehere@cisco.com} 
\IEEEcompsocitemizethanks{ \IEEEcompsocthanksitem $^\ddag$E-mail:sbhyri@cisco.com}
}
%%%%%%%%%%%%%%%%%%%%%%%%%%%%%%%%%%%%%%%%%%%%%%%%%%%%%%%%%%%%%%%%%%%%%%%%%%%%%%%%%%%%%%%%%%%%%%%%%%%%%%%%%%%%%%%%%%%%%%%%%%%%
\maketitle

\begin{abstract}
While the network traffic has seen exponential increase, the revenues have not maintained the same pace. New methods have to be explored to reduce this gap between traffic  and revenue. One such method is convergence in networking layers. In this work, we study the convergence of OTN and DWDM layer from a network planning perspective. We compare the costs of planning networks without and with convergence and show that the multilayer planning offers least cost for higher  traffic volumes. 
\end{abstract}


\IEEEpeerreviewmaketitle

\section{Introduction}

\IEEEPARstart{I}{nternet} traffic has witnessed enormous growth over the past decade driven by Data Centers and Cloud Applications. According to Cisco Visual Networking Index~\cite{Cisco}, the IP traffic is expected to grow three folds by 2019. 

The Core Optical Networks which form the backbone have also evolved from 10Gbps per wavelength to 100Gbps and beyond. With efficient Colorless, Directionless, Contentionless (CDC) and Flex Spectrum capable ROADM architectures, Optical Networks provide a touchless architecture for provisioning high bandwidth pipes on the fly.

While Optical Networks provide high bandwidth pipes, the client traffic is still of low bitrate. It is necessary to groom the low data rate client traffic to efficiently use the high capacity provided by the DWDM layer. The most effective solution for solving this problem is to have an Optical Transport Network (OTN) over DWDM technology~\cite{Nokia}.

Networks built with integrated OTN switching bring the economic and operational benefits of virtualization to optical networking. OTN switching decouples the clients from the DWDM line interfaces allowing greater network efficiency by ensuring that the more costly DWDM links are running as hot as possible and no stranded bandwidth remains. A network of OTN switches takes this concept further, allowing traffic to be aggregated at intermediate nodes and directed
towards underutilized routes~\cite{Infonetics}.

Some recent works propose multilayer optimization of IP over DWDM and IP over OTN over DWDM 
networks to reduce the CAPEX of the network~\cite{Pedrola,Ding,Claunir,RauII,Katib}. Work presented in~\cite{Eleni} considers the IP over OTN over DWDM network optimization also taking energy efficiency into account. Work in~\cite{Katib} employs a 
multilayer routing strategy by creating virtual links in all layers.

This work proposes ways to reduce the CAPEX of a multilayer OTN over DWDM network architecture. The end to end services are provisioned over the multilayer OTN over DWDM network considering the functionalities of both OTN and DWDM technologies. In section II, we discuss a heuristic algorithm for cost reduction. In section III, the results for a 30 node mesh and the National Science Foundation Network (NSFNET) are presented followed by conclusion in section IV. 

\section{Network planning and optimization}
\begin{figure}[!htbp]
 \begin{center}
 \includegraphics[scale = 0.27]{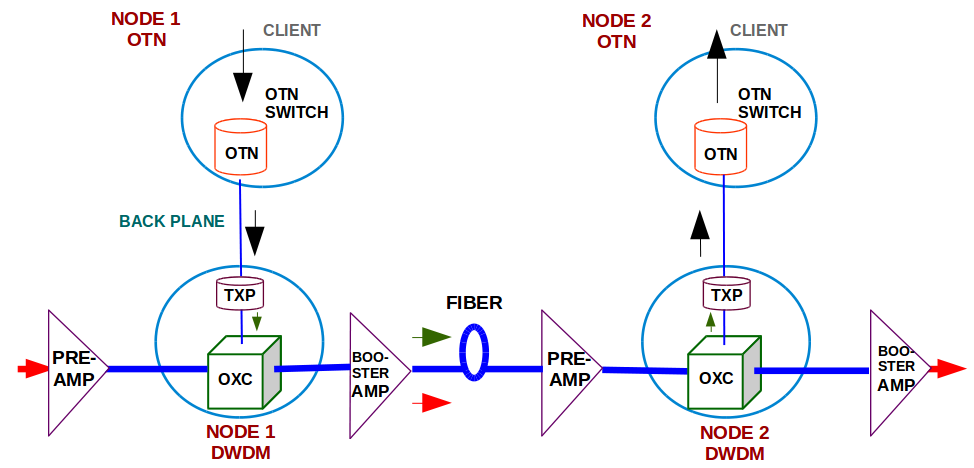}
\caption{A general architecture of OTN over DWDM network.}
\label{CostCalculation}
\end{center}
\end{figure}

OTN switching~\cite{ITU} allows multiple clients to be transparently bundled into uniform containers and sent over a single wavelength. OTN over DWDM system  can be implemented with  two different architectures (i) DWDM plus stand-alone OTN switch, where OTN switch and DWDM box are connected to each other by short reach optics, which drives to have  many back to back fiber interconnections (ii) Integrated OTN-DWDM system which eliminates the fiber interconnections~\cite{ofccnfoec}. In this work, an integrated approach is considered as shown in Fig.~\ref{CostCalculation}.

The assumptions are:
\begin{enumerate}[(a)]
\item At every node in the network, both the OTN and the DWDM equipment are present.
\item Client traffic aggregation at OTN box  are only in terms of ODU4.
\item Optical client and the OTN trunk are connected through the backplane.
\item DWDM node consists of a transponder, ROADM based OXC , booster amplifier and preamplifier.
\item Bandwidth variable transponders~\cite{Sambo, Svaluto} with data rates of 100Gbps and 200Gbps are considered.
 
\end{enumerate}
The OTN client traffic demands to be realized between source and destination OTN nodes are aggregated to form an ODU4 container. Aggregated ODU4 container at source OTN node traverses to the source DWDM node layer through backplane connectivity. Transponder in the DWDM layer performs an Electrical-to-Optical (E-O) conversion to the ITU-T wavelength. ROADM in the DWDM layer routes the wavelength towards the destination DWDM node. Booster and the preamplifiers present at the DWDM layer in the path compensate the node and fiber loss. If the path involves more than one span, ROADMs at the intermediate DWDM nodes choose to either express or drop the wavelength. The scenario of wavelength drop occurs when part of the ODU4 container has a different destination node. For example, consider a 3 node linear network, where there is an OTN request of 80G from Node 1 to Node 3 and an OTN request of 20G from Node 1 to Node 2. Both the OTN requests are aggregated into a single ODU4 container and onto a single wavelength. At Node 2, ROADM at the DWDM layer drops the wavelength. Transponder perform an O-E conversion and the data is sent to OTN layer. OTN switch drops the 20G meant for it and packs the 80G meant towards Node 3 into an ODU4 and sends it back to the DWDM layer after E-O conversion. ROADM at the DWDM layer routes the wavelength towards Node 3. ROADM at Node 3 drops the wavelength and the data is forwarded to OTN layer at Node 3 after an O-E conversion. ODU4 packet is demultiplexed by the OTN switch and the 80G meant for Node 3 is retrieved.

\subsection{Cost calculation model}
For a demand that traverses OTN over DWDM system, parameters that contribute to the cost are presented in Table I along with normalized cost value for each component. The cost incurred to satisfy an OTN over DWDM demand that calls for creation of new lightpaths is represented by equation~\ref{eq:1}.

\begin{dmath} \label{eq:1}
C_{New} =  d \times 2(C_{c} + C_{s} ) + n \times 2C_{t} + n \times C_{fc}.
\end{dmath}
where
\begin{dmath} \label{eq:2}
C_{fc}=\{(N-2) \times [2C_{oxc} +C_{ba} + C_{pa} ] + [2C_{oxc} + C_{ba} + C_{pa} ]\}/W,
\end{dmath}
d is the demand volume in Gbps, n is the number of wavelengths required to satisfy the demand, N is the number of DWDM nodes crossed, W is the total wavelengths supported in fiber, which is taken as 80 in this work, $C_{fc}$ is the cost of one wavelength traversing through Optical Crossconnect (OXC), Preamplifier and Booster amplifier. Since we assume back plane connectivity, $C_{otnt}$ and $C_{oc}$ are zero.

The cost incurred to satisfy an OTN over DWDM demand that is accommodated into existing lightpaths is given by equation~\ref{eq:3}.
\begin{dmath} \label{eq:3}
C_{Virtual} = 2 d \times  C_{c} + d \times C_{s} \times N.
\end{dmath}
\begin{table}[!htbp]
\begin{center}
\begin{tabular}{ | c | c | c | }
\hline
Abbreviation  &Parameter	                        & Normalized cost	\\  \hline \hline
$C_c$         &OTN client cost (per Gbps) 	        &	0.1	\\  \hline
$C_s$         &OTN switching cost (per Gbps)            &	0.01	\\  \hline
$C_{otnt}$    &OTN trunk cost                           &	0	\\  \hline
$C_{oc}$      &Optical client cost	                &	0	\\  \hline
$C_{t}$      &Transponder cost                  	&	40	\\  \hline
$C_{ba}$     & Booster Amplifier Cost (per unit)        &	20	\\  \hline
$C_{pa}$      &Pre-Amplifier Cost (per unit) 	        &	20	\\  \hline
$C_{oxc}$    &Cost of Optical Cross Connect (per unit)    &	30	\\  \hline

\end{tabular}
\caption{ Normalized cost associated with each of the resources used.}
\label{CostTable}
\end{center}
\end{table}
Given the input network topology and the traffic, the goal is to minimize the number of DWDM interfaces. This is achieved by efficiently using the switching and  grooming capability of  the  OTN  layer.  The optimization is compared with the transparent and the opaque switching~\cite{Ramamurthy} in network.

\subsection{The Heuristic Algorithm} 
\begin{figure}[!htbp]
\begin{center}
 \includegraphics[scale=0.65]{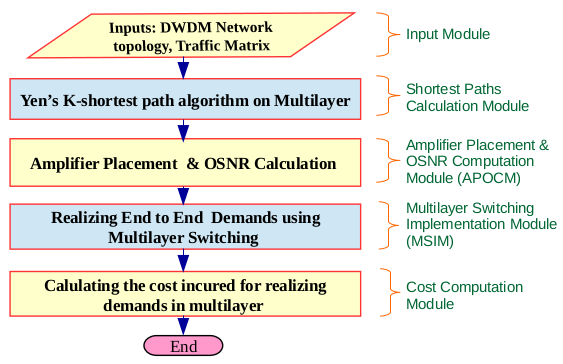}
\caption{Flowchart for the implementation of heuristic algorithm.}
\label{FlowChart}
\end{center}
\end{figure}

The proposed algorithm consists of five modules. Each module having a specific
functionality as shown in Fig.~\ref{FlowChart}.

The input module consists of a network topology which contains the set of fiber-connected nodes at Layer 0 
(DWDM) and the input traffic matrix. The traffic matrix contains the demands between OTN nodes. Every OTN node 
is connected to a DWDM node. The multilayer topology consists of one to one connection 
of OTN to DWDM node and fiber links interconnecting the DWDM nodes. There are no logical connections in 
the OTN layer at this stage. Logical connections are created dynamically by MSIM when lightpaths
are created.

The second module is the Shortest Paths Calculation Module which implements the Yen’s K shortest paths 
algorithm~\cite{Yen} on the DWDM topology using inputs from the first module to obtain K shortest paths between source and destination nodes. 

The third module is Amplifier Placement and OSNR Computation Module (APOCM). 
The amplifiers used are either EDFA24 or EDFA17~\cite{amplifier} 
at each DWDM node based on the architecture of DWDM node and parameters like degree of the node,
span loss and lump losses within the node. For simplicity, other nonlinear impairments have been ignored.
K shortest paths for each source and destination pair from the second module are arranged in decreasing
value of OSNR computed for each path~\cite{OSNR}.

\begin{figure}[!htbp]
\begin{center}
 \includegraphics[scale = 0.64]{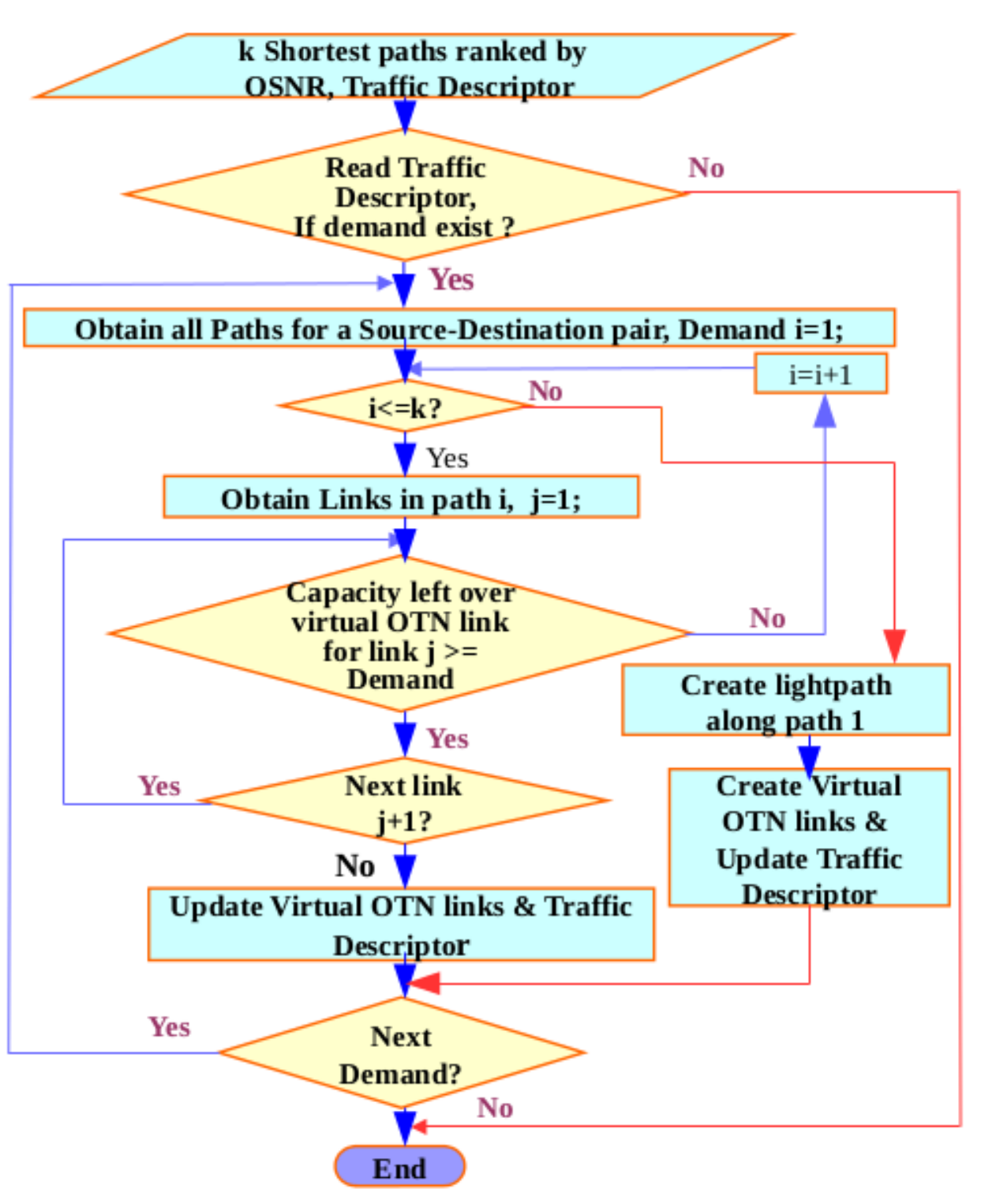}
\caption{MSIM flow chart.}
\label{MSIM}
\end{center}
\end{figure}

The fourth module is the Multilayer Switching Implementation Module (MSIM) as shown in Fig.~\ref{MSIM}. This module 
realizes the end to end traffic demands on OTN over DWDM network. From the traffic matrix, one demand is taken 
at a time and routed over the topology. At the beginning there are no links in the OTN layer, so initial traffic
is routed in the DWDM layer. Every time a lightpath is created in the DWDM layer, a corresponding link is created
in the OTN layer between the source and destination nodes of lightpath. The leftover capacity of the light path is
associated with the link. Subsequently, for each route in the k-paths set, existing OTN links with partial or complete
overlap with the route and with leftover capacity more than demand volume are mapped to the demand. The demand is then
routed over the existing OTN links without the need to create new light path. Cost is computed for each route based on
equation~\ref{eq:1} and~\ref{eq:3} and the one with minimum cost is finalized. It is observed in simulations that starting with 
the traffic demands between adjacent nodes yields better solutions in most cases as this leads to creation of OTN 
links with shortest light paths between add-drop nodes. The traffic demands between non-adjacent nodes are then routed
in decreasing order of their volume. Subsequently the leftover capacities in the OTN virtual links are updated. The
multilayer switching utilizes the left over capacity of lightpaths efficiently. This reduces the number of wavelengths
used which in turn reduces the number of transponders. 

\section{Results and discussion}
The two network topologies, the 30 node mesh (Fig.~\ref{30Node}) and the NSFNET (Fig.~\ref{NSFNET})
have been used to test the heuristic algorithm.
We consider the performance of an opaque optical network and transparent optical network for 
comparison of the proposed heuristic algorithm. The opaque network is the one where all traffic undergoes optical to 
electrical to optical (O-E-O) conversion at every hop. This ensures that the lightpaths are efficiently packed.
Transparent network is the one where end to end lightpaths are created for all demands. In transparent
network, traffic grooming is possible only at the end points. 

\subsection{Analysis of 30-node mesh network }

\begin{figure}[!htbp]
\begin{center}
 \includegraphics[scale = 0.37]{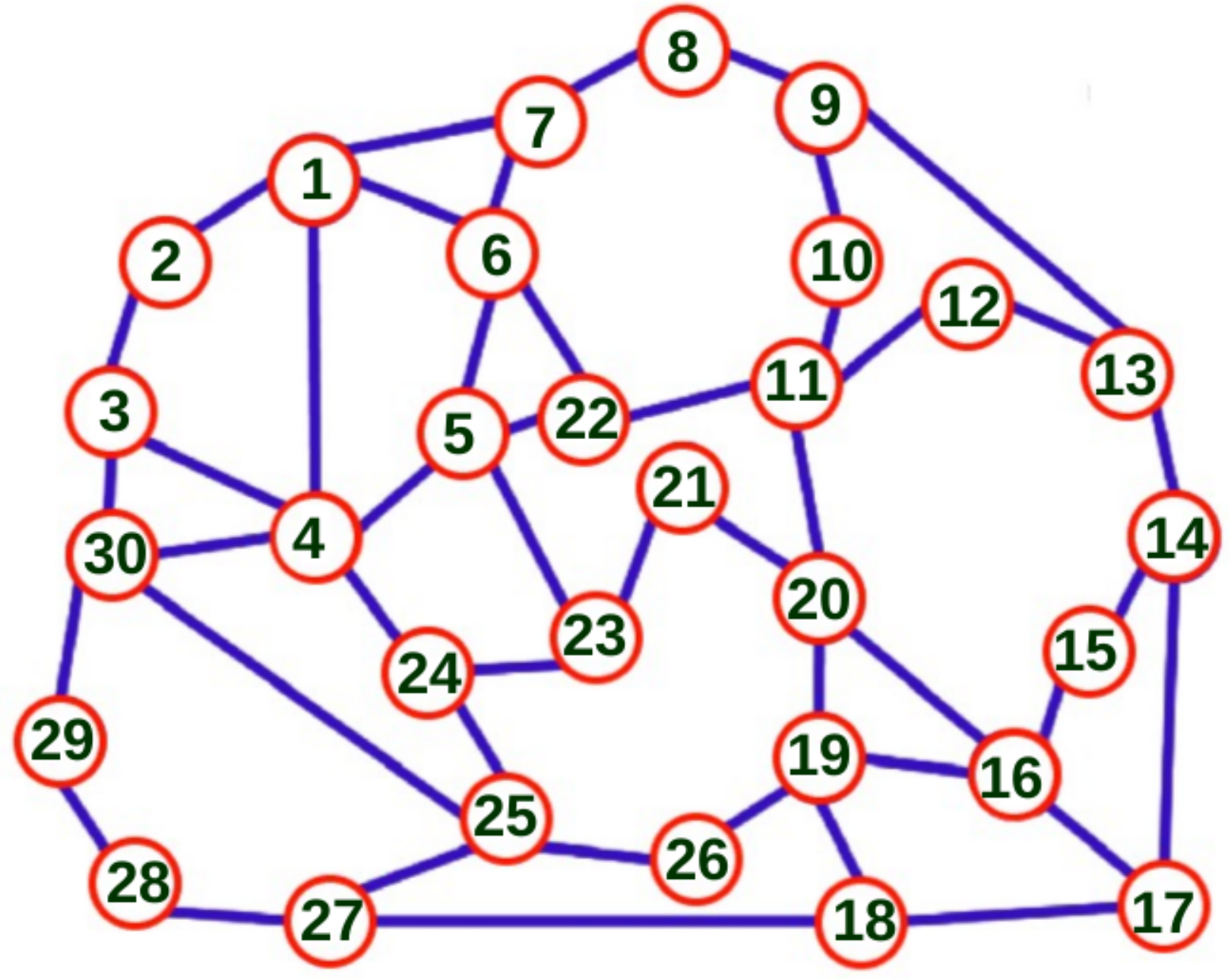}
\caption{Topology of the thirty node mesh network.}
\label{30Node}
\end{center}
\end{figure}

The 30 node mesh network~\cite{Sai} is shown in Fig.~\ref{30Node}. The link distance in the 30 node mesh
network is taken to be same for all the links. Traffic is assumed to be symmetric and it is incremented by 15\% in each phase to account for the traffic growth. Network topology along with the traffic matrix is input to the heuristic algorithm. Upto ten (k=10) shortest paths are found between the source and the destination node. MSIM takes the input of all the OSNR ranked shortest paths and after each demand is realized the capacity is updated on all links, both physical and logical.
 
For example, the OSNR ranked shortest paths from source 18 to destination 20, where the demand
to be realized, 40 Gbps, is the output from module three (APOCM) as shown in Table~\ref{1819}.
\begin{table}[!htbp]
\begin{center}
\begin{tabular}{ | c | c | c | c | c | }
\hline 
Source	& \shortstack{ Desti- \\ nation} &	Paths	                        &    OSNR (dB)  &     $C_{New}$	\\  \hline \hline
 18	&	20	                 &	18-19-20	                &	19.96 &	91.30\\  \hline
 18	&	20	                 &	18-17-16-20                     &	19.23 &	92.55\\  \hline
 18	&	20                     	&	18-19-16-20              	&	19.21 &	92.55\\  \hline
 18	&	20	                 &	18-17-16-19-20 	                &	19.10 &	98.80\\  \hline
 18	&	20	                 &	18-27-25-26-19-20	        &	18.36 &	95.05\\  \hline
 18	&	20	                 &	18-27-25-24-23-21-20      	&	17.43 &	96.30\\  \hline
 18	&	20                     	&	18-27-25-24-23-5-22-11-20 	&	16.82 &	98.80\\  \hline
 18	&	20	                &	18-17-16-15-14-13-12-11-20	&	16.73 &	98.80\\  \hline
 18	&	20	                &	18-27-25-24-4-5-22-11-20	&	16.50 &	98.80\\  \hline
 18	&	20	                &	18-27-28-29-30-4-5-23-21-20	&	16.31 &	100.05\\  \hline

\end{tabular}
\caption{OSNR ranked paths for source-destination pair 18-20.}
\label{1819}
\end{center}

\end{table}

MSIM checks if there is an existing virtual link between the two nodes with sufficient capacity to satisfy the demand,
and if so, it is routed over the virtual links. As seen in table 3, for the above example path 18-17-16-20 has
required capacity and hence it is chosen for the demand  MSIM finds the path with each link capacity greater than 
the demand to be realized. Path chosen for the above example is 18-17-16-20 as from Table~\ref{1819} and the capacity 
that is available in each of the links is shown in Table~\ref{Left}.

\begin{table}[!htbp]
\begin{center} 
\begin{tabular}{ | c | c | c |  }
\hline
Source link	&	Destination link	&	Capacity  in virtual link (Gbps)	\\  \hline \hline
18	        &	17	&	80	\\  \hline
17	        &	16	&	40	\\  \hline
16	        &	20	&	50	\\  \hline

\end{tabular}
\caption{Capacity left over in OTN virtual link for source 18 and destination 20 before realizing the demand.}
\label{Left}
\end{center}
\end{table}

\begin{table}[!htbp]
\begin{center}
\begin{tabular}{ | c | c | c | c | }
\hline 
Source	& Destination &	Paths &           $C_{Virtual}$	\\  \hline \hline
18	&	20 &	18-17-16-20   & 9.60            \\  \hline
\end{tabular}
\caption{Cost incurred in realizing a demand from 18-20 using virtual OTN path.}
\label{Virtual}   
\end{center}
\end{table}

The number of wavelengths needed to realize the demand is zero. Thus the number of transponders
needed is zero as the demand is realized using virtual leftover capacities which were created by
previous lightpath in the DWDM layer. If the demand is realized by setting up a new lightpath
then the cost incurred is 92.55 as in Table~\ref{1819}. Now since the demand is realized by the
virtual OTN links the cost incurred is 9.6 as in Table~\ref{Virtual}. Thus MSIM optimizes the
cost incurred to realize the demand choosing feasible OSNR paths.   

\begin{figure}[!htbp]
\begin{center}
 \includegraphics[scale=0.52]{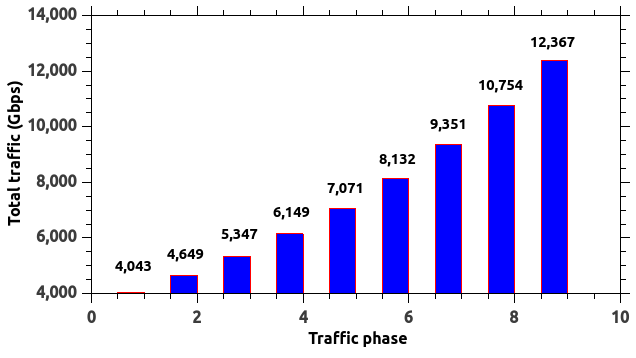}
\caption{Net traffic in different phases for analyzing 30 node mesh network with 100Gbps transponders.}
\label{30Node100GPhase}
\end{center}
\end{figure}

\begin{figure}[!htbp]
\begin{center}
 \includegraphics[scale=0.54]{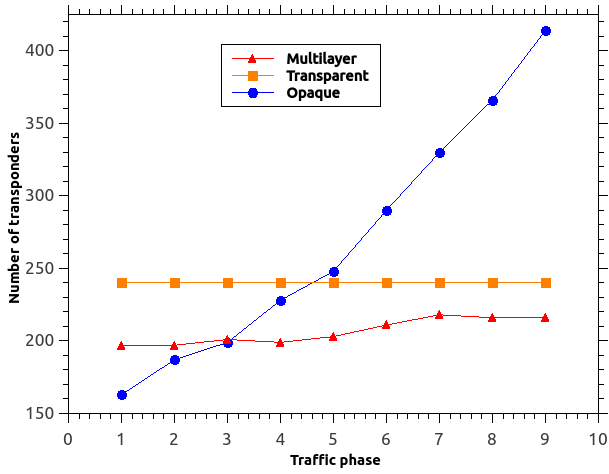}
\caption{Plot of traffic phase vs. number of 100Gbps transponders for the 30 node mesh network.}
\label{30Node100G}
\end{center}
\end{figure}

Bandwidth variable transponders capable of supporting 100Gbps and 200Gbps per 
wavelength are used in the optical layer. For each traffic phase as in Fig.~\ref{30Node100GPhase},
all the demands are optimally realized by the heuristic algorithm with 100Gbps transponders.
This is compared with opaque and transparent network scenario in Fig.~\ref{30Node100G}. 

\begin{figure}[!htbp]
\begin{center}
 \includegraphics[scale=0.5]{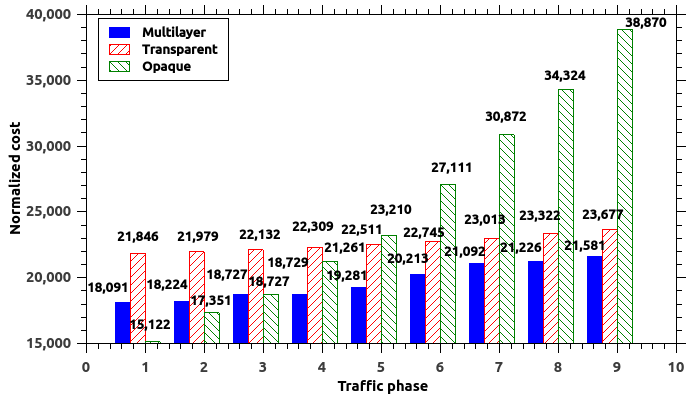}
\caption{Plot of traffic phases vs.  normalized cost for various switching with 100Gbps transponders in the 30 node mesh network.}
\label{Cost100G30}
\end{center}
\end{figure}

We infer from Fig.~\ref{30Node100G} that in low traffic scenario, opaque network requires least number of transponders.
But as the traffic grows the O-E-O conversion becomes expensive requiring more transponders and a bypass at optical 
layer is a better option. On the other hand, the transparent network results in inefficient use of transponders at low traffic due to low packing of 
the wavelength capacity. It is evident from the plot that using multilayer optimization adopts the benefits of both,
leading to optimized number of transponders. The normalized cost incurred for different switching using 100Gbps 
transponders is in Fig.~\ref{Cost100G30}, which shows that in low traffic scenario, opaque network costs the least
and as the traffic grows the multilayer switching is more economical.

\begin{figure}[!htbp]
\begin{center}
 \includegraphics[scale=0.5]{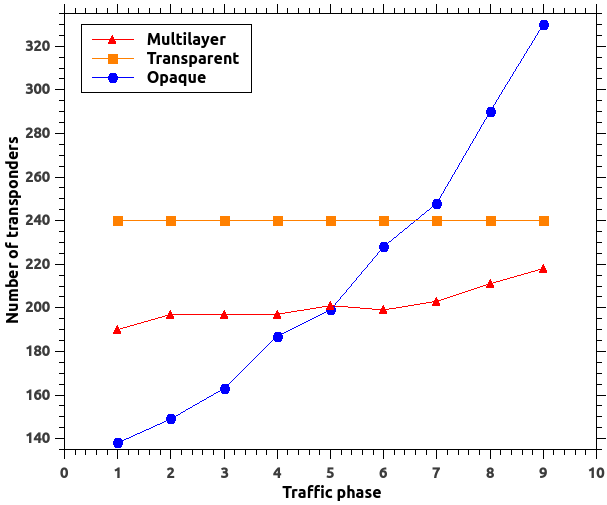}
\caption{Plot of traffic phases vs. number of 200Gbps transponders for the 30 node mesh network.}
\label{30Node200G}
\end{center}
\end{figure}

A similar analysis is repeated for 200Gbps transponders with an incremental traffic phase. 
The Fig.~\ref{30Node200G} compares the number of transponders in the optical layer from the heuristic algorithm 
with the opaque and the transparent network scenario. 

\begin{figure}[!htbp]
\begin{center}
 \includegraphics[scale=0.48]{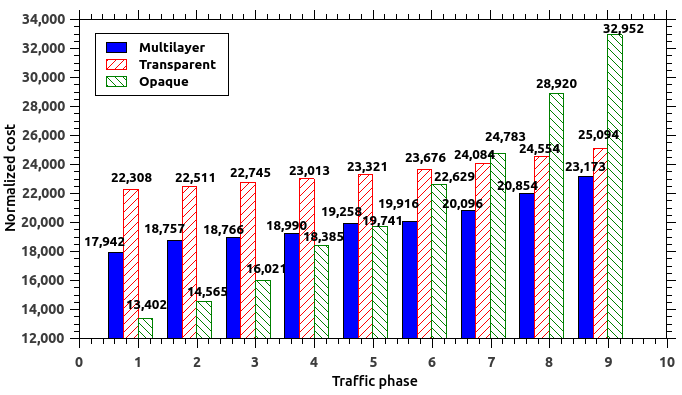}
\caption{Plot of traffic phases vs.  normalized cost for various switching with 200Gbps transponders in the 30 node mesh network.}
\label{Cost200G30}
\end{center}
\end{figure}

The cost incurred for different switching using 200Gbps transponders
is shown in Fig.~\ref{Cost200G30}. The graphs for 200Gbps transponders in 30 node mesh network again shows that the opaque 
switching requires the least number of transponders in low traffic scenario and as the traffic in the network grows 
the multilayer switching is more economical.

\subsection{Analysis of the NSFNET}
\begin{figure}[!htbp]
\begin{center}
 \includegraphics[scale = 0.42]{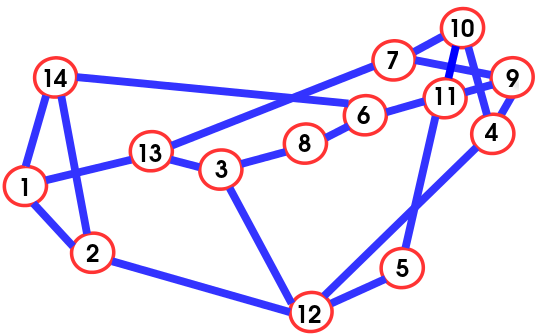}
\caption{The NSFNET topology.}
\label{NSFNET}
\end{center}
\end{figure}

\begin{figure}[!htbp]
\begin{center}
 \includegraphics[scale=0.55]{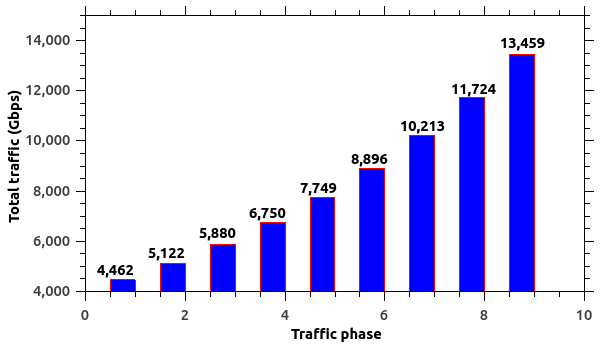}
\caption{Net traffic in different phases for analyzing NSFNET with 100Gbps transponders.}
\label{NSF100GPhase}
\end{center}
\end{figure}

\begin{figure}[!htbp]
\begin{center}
 \includegraphics[scale=0.53]{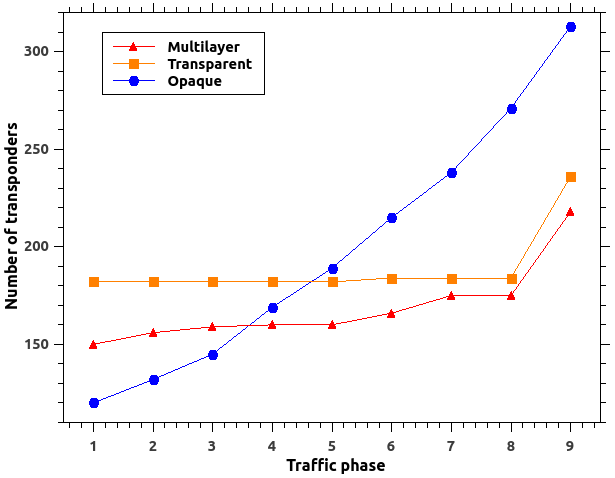}
\caption{Plot of traffic phases vs. number of transponders required with 100Gbps transponder in the NSFNET.}
\label{NSF100G}
\end{center}
\end{figure}

\begin{figure}[!htbp]
\begin{center}
 \includegraphics[scale=0.5]{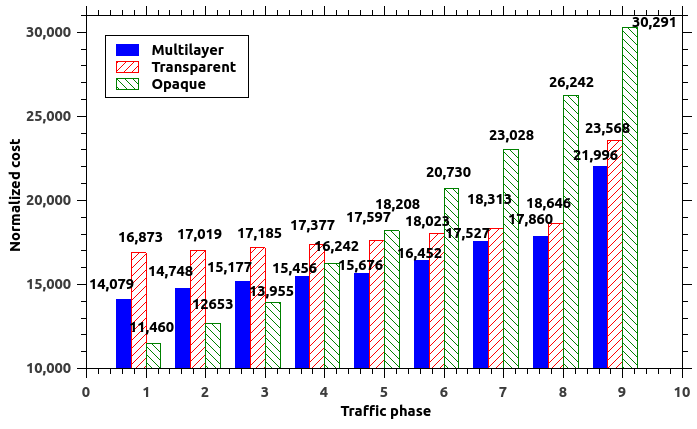}
\caption{Plot of traffic phases vs.  normalized cost for various switching with 100Gbps transponders in the NSFNET.}
\label{Cost100GNSF}
\end{center}
\end{figure}

The National Science Foundation Network (NSFNET) is a 14-node topology with 21 links~\cite{Katib}.
The NSFNET topology is shown in Fig.~\ref{NSFNET}. 
The heuristic algorithm is implemented on this network. Since the link distances here are not same for 
all the links, the choice of preamplifier used is dependent on the link distance. The algorithm gives the details of total traffic realized and the 
total number of transponders required. Fig.~\ref{NSF100G} shows the number of 100Gbps transponders required
to realize the network satisfying its traffic demands by implementing multilayer switching,
transparent and opaque switching respectively for traffic phase as in Fig.~\ref{NSF100GPhase}.  
The normalized cost incurred for different switching using 100Gbps transponders is shown in Fig.~\ref{Cost100GNSF}.

\begin{figure}[!htbp]
\begin{center}
 \includegraphics[scale=0.543]{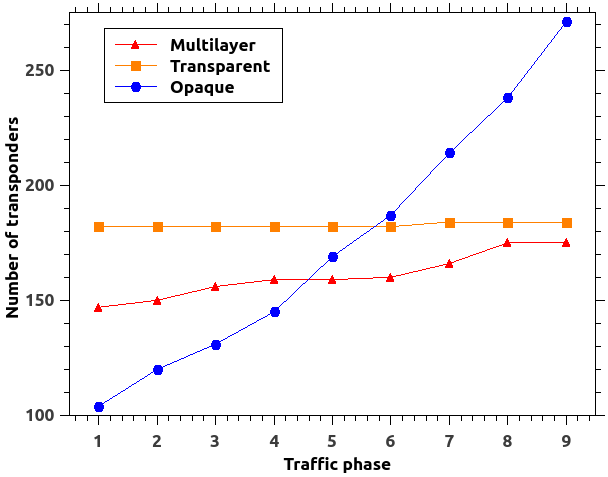}
\caption{Plot of traffic phases vs. number of transponders with 200Gbps transponders in the NSFNET.}
\label{NSF200G}
\end{center}
\end{figure}

\begin{figure}[!htbp]
\begin{center}
 \includegraphics[scale=0.48]{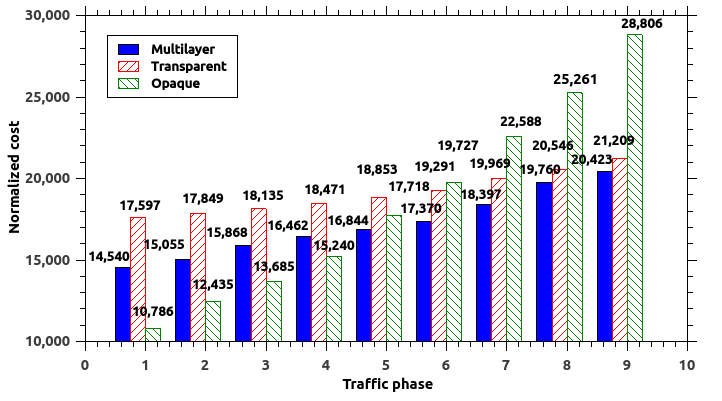}
\caption{Plot of traffic phases vs. normalized cost for various switching with 200Gbps transponders in the NSFNET.}
\label{Cost200GNSF}
\end{center}
\end{figure}

A similar analysis is repeated for 200Gbps transponders with an incremental traffic phase. 
A comparison of the number of transponders in the optical layer with opaque and transparent network scenario is shown
in Fig.~\ref{NSF200G}. The normalized cost incurred for different switching using 200Gbps transponders is shown in Fig.~\ref{Cost200GNSF}.

We see from the plots for the NSFNET, as inferred from 30 node mesh network, that opaque switching requires the least number
of transponders in low traffic scenario and multilayer switching is more economical as the traffic in the network grows. It
is evident that using multilayer optimization adopts the benefits of both,
leading to optimized number of transponders. From Figs.~\ref{30Node100G}, \ref{30Node200G}, \ref{NSF100G} and \ref{NSF200G}
we see that the optimization becomes more relevant when the traffic is increased in the optical 
layer. This is very relevant given that while the client traffic data rate still remains
sub-10G, the wavelength capacity has increased to 200G and beyond in the optical layer.

\section{Conclusion}
While OTN protocol implemented over DWDM offers many management and control plane benefits,
the overall network CAPEX can be minimized in the design phase by planning both layers simultaneously.
The multilayer switching finds a middle path between opaque and transparent network over all traffic volumes. 
The multilayer switching introduced solves two issues: bandwidth management and underutilization of transponders.
This approach can also be applied during the periods of the low traffic in the day when it can be switched and concentrated on a few
links while shutting down the other transponders and reducing the power consumption of the network. 

The OSNR for the path gives the optical feasibility and reliability of the end to end
lightpath. This approach gives reliable end to end service since it takes
the lightpaths that are optically feasible. The demands realized over virtual OTN links are
also reliable, as it uses an existing lightpath.

The heuristic algorithm proposed using multilayer switching converges the layers involved and 
helps in efficient planning of the network leading to the optimum utilization of the 
resources along with providing reliable service. Multilayer OTN over DWDM switching
is best  suited for metro networks, where the demand relative to the network is high. 

\appendices
\bibliography{IEEEAnts}

\bibliographystyle{./IEEEtran}
\end{document}